\documentclass{article}

\usepackage{ucs}
\usepackage[utf8x]{inputenc}
\usepackage[LGR,T2A,T1]{autofe}

\usepackage{graphicx}
\usepackage{amsmath}
\usepackage{url}

\usepackage{stringenc}
\usepackage[filenameencoding=utf8x]{grffile}

\usepackage{listings}
\begin{document}

\title{Snafu: Function-as-a-Service (FaaS) Runtime Design and Implementation}
\author{Josef Spillner\\
Zurich University of Applied Sciences, School of Engineering\\
Service Prototyping Lab (blog.zhaw.ch/icclab/)\\
8401 Winterthur, Switzerland\\
josef.spillner@zhaw.ch}

\maketitle

\begin{abstract}
Snafu, or Snake Functions, is a modular system to host, execute and manage language-level functions offered as stateless (micro-)services
to diverse external triggers.
The system interfaces resemble those of commercial FaaS providers but its implementation provides
distinct features which make it overall useful to research on FaaS and prototyping of FaaS-based applications.
This paper argues about the system motivation in the presence of already existing alternatives, its design and architecture,
the open source implementation and collected metrics which characterise the system.
\end{abstract}

\section{Introduction}

The construction of software applications went through several historical trends.
Early on, software programs were subdivided into functions, later into classes with methods,
modules, higher-level components with well-defined interfaces and eventually into uniform service
interfaces. As services grew and became monolithic, the trend was reversed, microservices became
popular \cite{microservicesscale}, and eventually fine-granular single functions are now again the unit of choice for
encapsulating service functionality in many domains. This trend is leading to cloud applications
with fine-grained billing and seemingly serverless hosting \cite{chatbot} and to improved computing
abilities for mobile and connected devices \cite{dripcast} and scientific workflows \cite{serverlessworkflows}.

Function-as-a-Service (FaaS) is therefore the technological concept to subdivide software applications
into functions and to invoke these functions through network protocols with explicitly specified or
externally triggered messages. The functions are deployed as FaaS units, encompassing the callable
functions along with their dependencies, on dedicated FaaS runtimes.
The runtimes are platform-level host services whose tasks are the blazingly fast allocation of
resources for the function execution and the precise accounting of the associated duration and processing load.
Functions may be executed in-process or externally through interpreters, containers or other isolation layers.
They are generally stateless, losing all local state after termination, and require bindings to external
stateful services such as key-value stores, databases, file or blob storage for any persistence.

This paper positions Snafu as runtime system implementation to host in-process Python functions and external functions
programmed in other languages within a FaaS environment.
Snafu is flexible, low-effort and competitive with regards to similar runtimes and resolves the
need for a FaaS runtime geared towards not only cloud providers but also cloud application engineers.
The text is structured as follows.
All of the next five sections start with a question which the section is answering to.
The sections describe the motivation, design, architecture, implementation and experimental evaluation of Snafu.
A last section concludes with an outlook on future work.

\section{Motivation}

Why spend time and effort on Snafu when there are already plenty of FaaS hosts for Python functions and multi-language
functions available?
Table \ref{tab:pythonfaas} compares currently available open source and commercial FaaS runtimes which specifically
target Python software engineers or application providers, most of which are recent additions to the market within
the timeframe of 2015 to 2017. Only one service is commercially available on demand. Comparison tables for other
languages would look similar with the notable exception of JavaScript (Node.js) which sees more wide-spread support.

\begin{table}[htb]
\centering
\caption{FaaS runtimes targeting Python functions.\label{tab:pythonfaas}}
\begin{tabular}{|l|l|} \hline
\textbf{Runtime}		& \textbf{Classification}		\\ \hline

AWS Lambda			& commercial service			\\ \hline
Docker-LambCI			& open source, reverse engineered	\\ \hline
OpenLambda \cite{serverless}	& open source, research prototype	\\ \hline
PyWren \cite{pywren}		& open source, AWS Lambda overlay	\\ \hline
Fission				& open source				\\ \hline
Kubeless			& open source, Kubernetes-integrated	\\ \hline
IronFunctions			& open source, AWS Lambda compatible	\\ \hline
\end{tabular}
\end{table}

Many of these runtimes target infrastructure-level deployment and are not suitable for quick prototyping
of applications or ad-hoc experiments.
The main motivation for a different design is therefore flexibility coupled with a reduced effort when getting starting with the system.

\begin{itemize}
\item Flexibility: AWS Lambda is known to have several operational limits of which only a subset can be configured or requested to be configured differently \cite{cloudcontrolplane}. In many scenarios, it is desirable to change limits or other details of the implementation which necessitates a fully controllable open source approach. The open source runtimes can be modified but often require code-level modifications for even basic extensions. The mapping from functions or methods in the programming scope to functions offered as a service in the FaaS scope is often incomplete.
\item Reduced effort: Several of the open source FaaS runtimes, both AWS Lambda-compatible and incompatible ones, require a non-trivial setup and integration with other runtime components and frameworks. Reduced effort means that basic tasks should work without any configuration to achieve first results quickly, and more complex tasks should involve minimal configuration. Awareness for this requirement is increasing in cloud research \cite{cloudcomplexity}.
\end{itemize}

From these explanation, concrete criteria are derived which allow for a comparative evaluation.
Criteria thus encompass: Must be available as open source (C1), must be deployed and operational for simple tests within
an attention span of ten minutes
assuming the required basic setup and dependency installation has been performed (C2), and must work in some configuration
without requiring other runtimes (C3). Furthermore, the runtime must consume source code and extract functions or methods
found within (C4) to allow for quick conversion of legacy codebases.
The motivation for a custom FaaS design is justified when it can be demonstrated that the existing runtimes do not fulfil these criteria.
Table \ref{tab:criteria} roughly calculates how the existing runtimes score with one point given for each fulfilled criterion.
The comparison shows that Docker-LambCI is mainly targeting Node.js whereas the Python part is not fully
implemented, making it impossible to run Python scripts. OpenLambda and IronFunctions require Docker containers as isolation layer,
and the former returns an empty string from the hello world function instead of the expected result, indicating a temporary bug.
PyWren requires the creation of custom Lambda roles.
Fission and Kubeless strictly require access to a running Kubernetes instance.
Most runtimes require explicit function selection and configuration instead of dynamically exporting any function.
This comparative feature calculation is not meant to be precise, but it gives an indication of the issues associated with current systems.
Half points (½) are given for features which come close but do not fully fulfil the criteria.

\begin{table}[htb]
\centering
\caption{FaaS runtimes targeting Python functions.\label{tab:criteria}}
\begin{tabular}{|l|l|l|l|l|l|} \hline
\textbf{Runtime}& \textbf{C1}	& \textbf{C2}	& \textbf{C3}	& \textbf{C4}	& \textbf{Sum}	\\ \hline

AWS Lambda	& 0		& 1		& 1		& 0		& 2	\\ \hline
Docker-LambCI	& 1		& 0		& 0		& 0		& 1	\\ \hline
OpenLambda	& 1		& ½		& 0		& 0		& 1½	\\ \hline
PyWren		& 1		& 0		& 0		& ½		& 1½	\\ \hline
Fission		& 1		& 0		& 0		& 0		& 1	\\ \hline
Kubeless	& 1		& 0		& 0		& 0		& 1	\\ \hline
IronFunctions	& 1		& 1		& 0		& 0		& 2	\\ \hline
\end{tabular}
\end{table}

None of the existing runtimes fulfil all criteria ($sum = 4$) which justifies either their improvement or a new
design and implementation. Snafu follows a new design to evaluate the ease of prototypical engineering
of an alternatively designed FaaS runtime.

\section{Design}

What are the design criteria to achieve a more flexible and effort-reduced FaaS host?
Compared to the existing approaches, Snafu is designed to execute functions written in its implementation language Python in-process
and functions in arbitrary languages through external processes.
This design does therefore not preclude the addition of languages executed through an isolation layer such
as spawned interpreters or containers. Furthermore, the design mandates a zero-configuration approach. The user
experience after downloading the host should be instant by offering runnable sample functions
and client functionality which works out of the box.
To foster adoption, a migration path from commercial FaaS services, which for Python functions currently
implies AWS Lambda, is integral to the design.
Files which contain more than one function can be configured in a way that each function is
exported so that code redundancy through multiple deployed function units is decreased.
Triggers should be extensible and a suitable number of useful triggers is part of the
software.

\begin{figure}[h]
\center
\includegraphics[width=0.596\columnwidth]{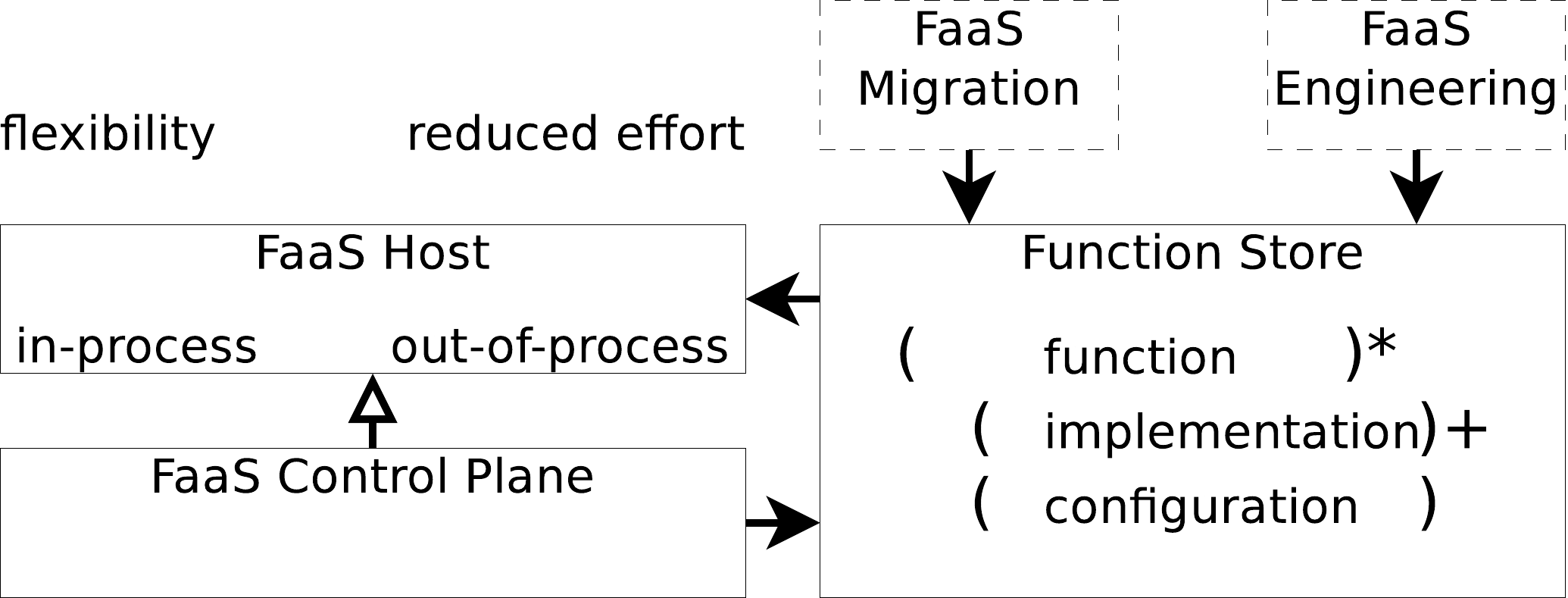}
\caption{Snafu design}
\label{fig:design}
\end{figure}

Fig. \ref{fig:design} shows the Snafu overall functional design.
Functions consisting of implementation and configuration are either engineered from scratch or migrated from an existing FaaS.
They are stored in a pool from which they can be executed with a host process which can be remote-controlled through a
control plane with function management service interfaces.

\section{Architecture}

Which software architecture is appropriate for Snafu, given the design guidelines and constraints?
As with the functions it supports, modularity and service-orientation are important architectural
properties of the system itself.

The architecture follows the design. Flexibility is achieved by an extensible system whose
functionality is partly contained in subsystems with pluggable components. Reduced effort
is achieved by a zero-configuration default setup as well as flexible per-instance, per-tenant
and per-function configuration options, in order of priority.

\begin{figure}[htb]
\center
\includegraphics[width=0.70\columnwidth]{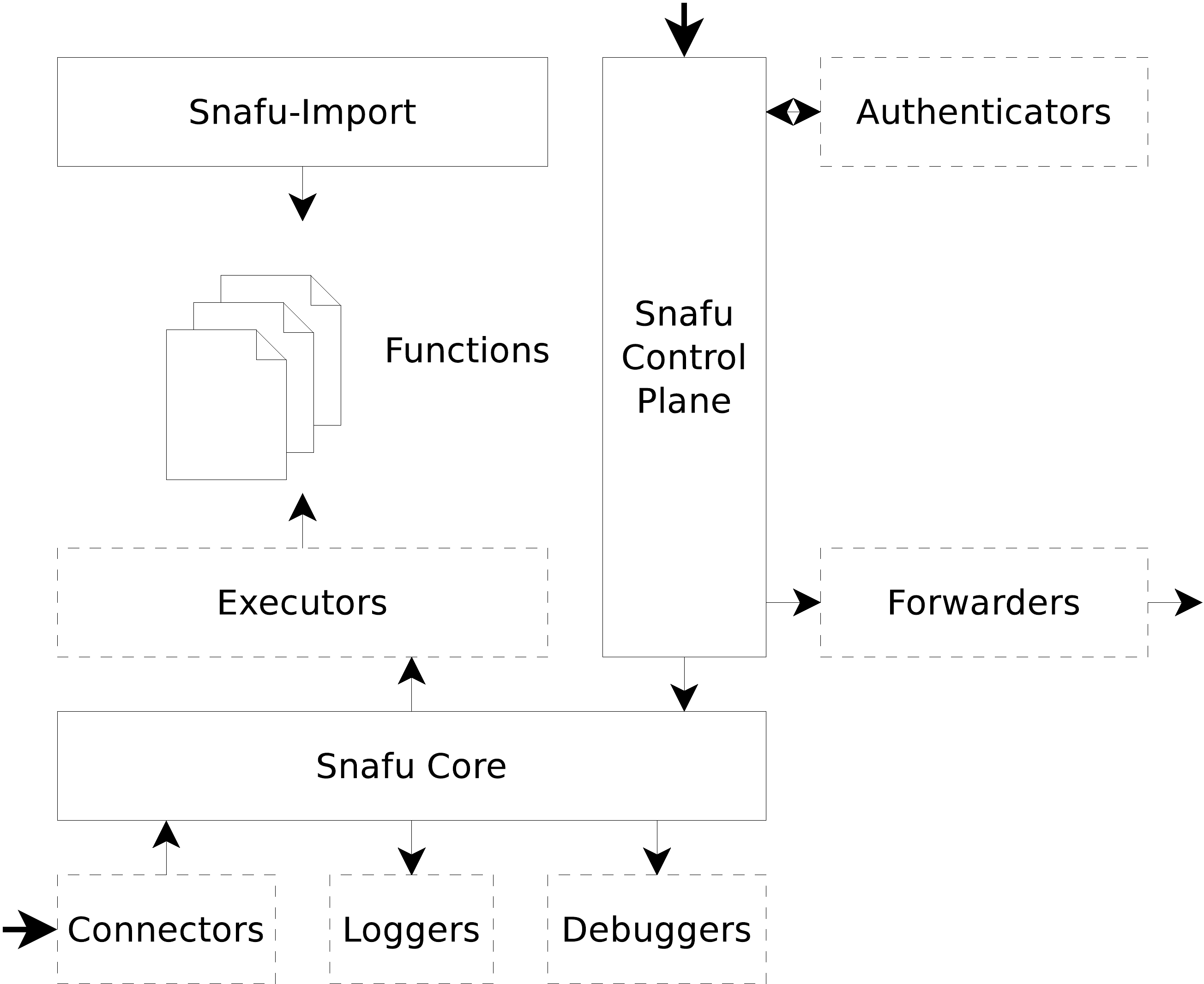}
\caption{Snafu architecture}
\label{fig:architecture}
\end{figure}

As shown in Fig. \ref{fig:architecture}, the two main
building blocks are the FaaS host core and the control plane. Both blocks are complemented
by six extensible subsystems for triggering, authenticating, logging, debugging, forwarding
and finally execution of functions. Support for Lambda functions is offered through a dedicated
import path which imports all functions from an AWS region in a batch process, as well as
through the compatible control plane which apart from Lambda also implements the necessary
service interfaces of AWS S3 and STS.

Functions may be invoked locally through the Snafu tool which interacts with humans through
a command-line interface. Requests are also received by the control plane which, depending
on the configuration, authenticates the request, decides about routing it to a pre-configured
static or dynamically launched secondary instance of Snafu (e.g. for per-tenant isolation or load
balancing) or processing it internally. In the latter case, the request, execution and response
of the function are properly logged and amended with debugging information. The execution is performed
in-process or out-of-process.
A third channel for invocation are the connectors which make functions available through triggers
such as web services, message queue subscriptions and timers.

\section{Implementation}

How is Snafu implemented in order to maintain the design criteria of flexibility and reduced
effort?
The implementation consists of a set of Python 3 executables and library modules as well as parts implemented in other languages which
cover the entire functionality shown in the architecture (Fig. \ref{fig:architecture}).
The executables are the \texttt{snafu} command-line tool as function host process, \texttt{snafu-import} to import already
deployed functions in AWS Lambda, and \texttt{snafu-control} to run a Lambda-compatible control plane.
The executables interact with the filesystem to read and write function files, and with various
triggers to receive events. Five triggers are implemented: web (HTTP invocations), messaging (AMQP
messages), filesystem (file change notifications through inotify), cli (interactive command-line input),
and cron (scheduled regular invocations).

The implementation possesses both in-process and out-of-process executors. The former group is limited to Python 3
functions which matches the implementation language.
The choice of Python 3 as implementation language renders Snafu in-process function execution initially incompatible with
AWS Lambda which only supports Python 2.7.
To mitigate any friction, Snafu can be instructed to automatically upgrade all Python 2 code to Python 3 either upon
explicit import or upon the creation of new functions, using Python's \texttt{2to3} utility.
Furthermore, among the implemented external executors one supports Python 2 for increased compatibility due to the
effective but still heuristic \texttt{2to3} conversion algorithm. Another external executor runs Java methods as
functions which is otherwise a unique feature of AWS Lambda among the commercial FaaS providers.
In total, five executors with different language and isolation capabilities are available.

For scalability and multi-tenancy, forwarders are available which redirect requests to other instances of Snafu which
are static or created dynamically per tenant. Tenants are authenticated against an accounts list by any number of
authenticators.
Compatibility with AWS Lambda is achieved by the availability of an AWS4 signature verification among the authenticators.

\begin{figure}[htb]
\center
\includegraphics[width=0.85\columnwidth]{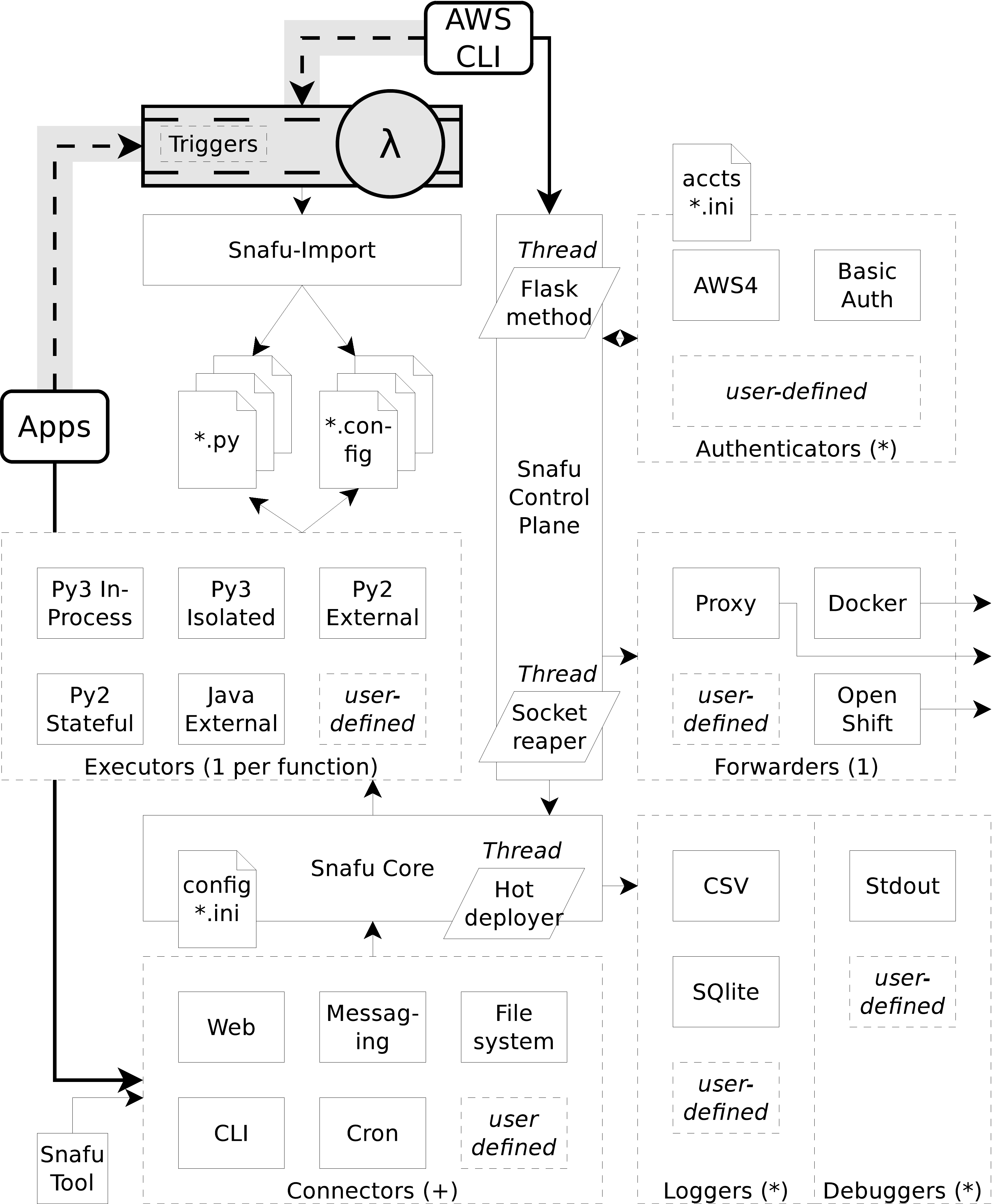}
\caption{Snafu implementation}
\label{fig:implementation}
\end{figure}

The complete implementation architecture is shown in Fig. \ref{fig:implementation}. All subsystems can be extended by
placing additional Python files adhering to their function signature constraints into the respective subsystem directories.
The parts marked in grey belong to existing commercial FaaS providers and are not required for operation.

Multi-threading is used on three occasions. Each remote request received by the control plane creates a processing thread
to ensure parallel processing of multiple requests. Furthermore, if requested, one static thread monitors the sockets
of all incoming connections (reaper thread) and another one monitors the filesystem for hot-deployed functions (hot deployer
thread).

The entire codebase of Snafu consists of around 1300 lines of Python 3 code and smaller footprints for Python 2 and Java 8 parsing
and execution. To give an impression of how the tool is used in versatile ways: \texttt{snafu -x helloworld hello.py} would execute
the function $helloworld$ from the specified source file. Any function argument not already known is queried interactively from the
user if possible. Depending on the connector and calling convention, the parameters \texttt{event} and \texttt{context} may be preset
per convention among commercial FaaS implementations. The invocation of \texttt{snafu-control -e docker -a aws} would run each
function after authentication within a per-tenant Docker container.

\section{Metrics}

\begin{figure}[h]
\center
\includegraphics[width=0.443\columnwidth]{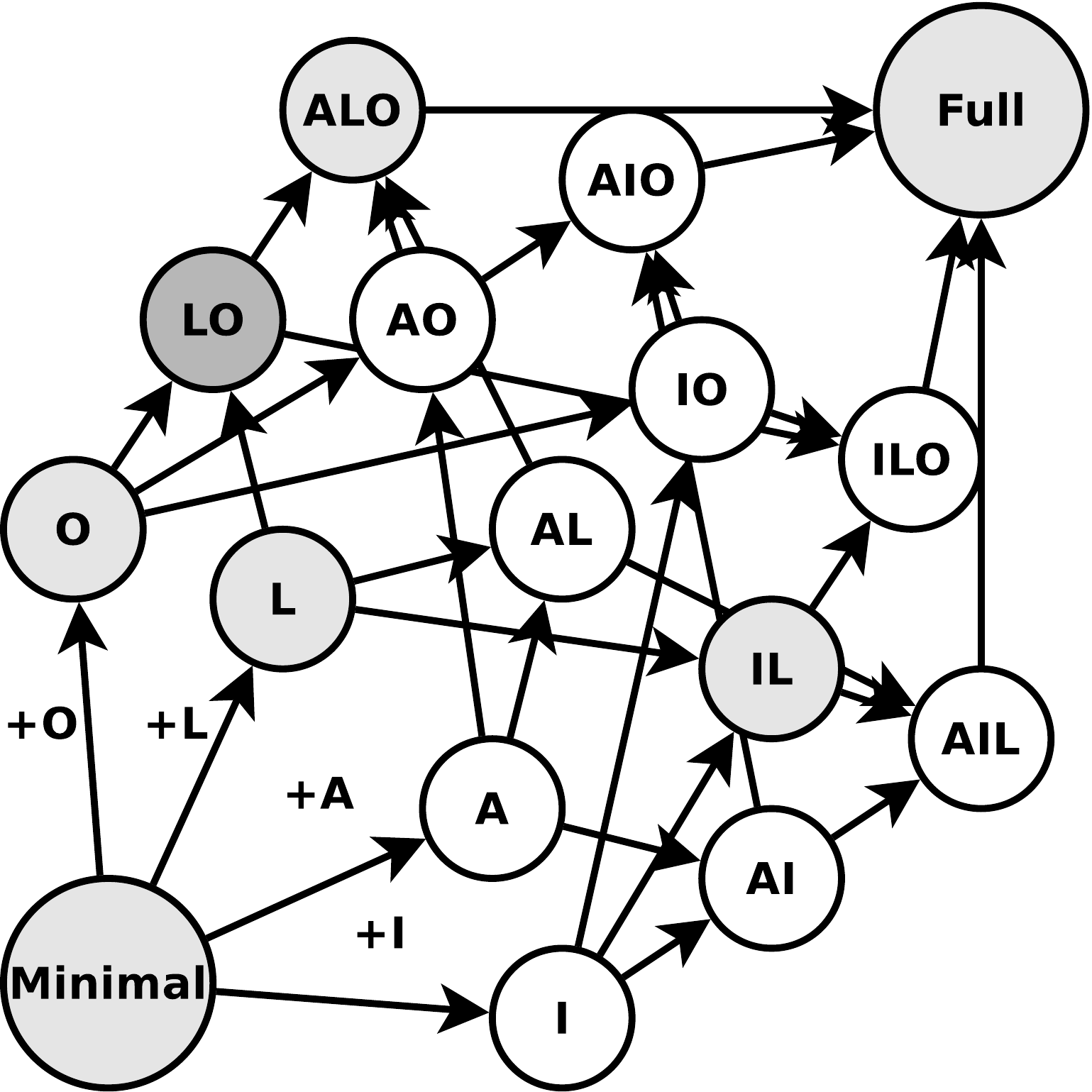}
\caption{Snafu execution configuration space: O = output for debugging, L = logging, A = authentication, I = isolation}
\label{fig:confspace}
\end{figure}

How does Snafu perform and scale? This section reports about an experimental evaluation of the software
implementation according to various configurations. The configuration space for the internal (in-process) Python function
execution is shown in Fig. \ref{fig:confspace}. The light grey-highlighted measured configurations are complemented with
external (out-of-process) executors and forwarders, and the dark grey configuration represents the implementation default.

In the experiments, $O$ is represented by debug messages written to the process standard output.
$L$ is represented by CSV logs, $A$ by the AWS4 digital signature verification and $I$
by the stateless function isolation.

The configurations were measured with concrete Snafu subsystem configurations. All local experiments
were performed on a contemporary notebook with a quad-core i7-5600U CPU clocked at 2.60GHz and
16 GB RAM.

\subsection{Performance}

\begin{figure}[h]
\center
\includegraphics[width=0.454\columnwidth]{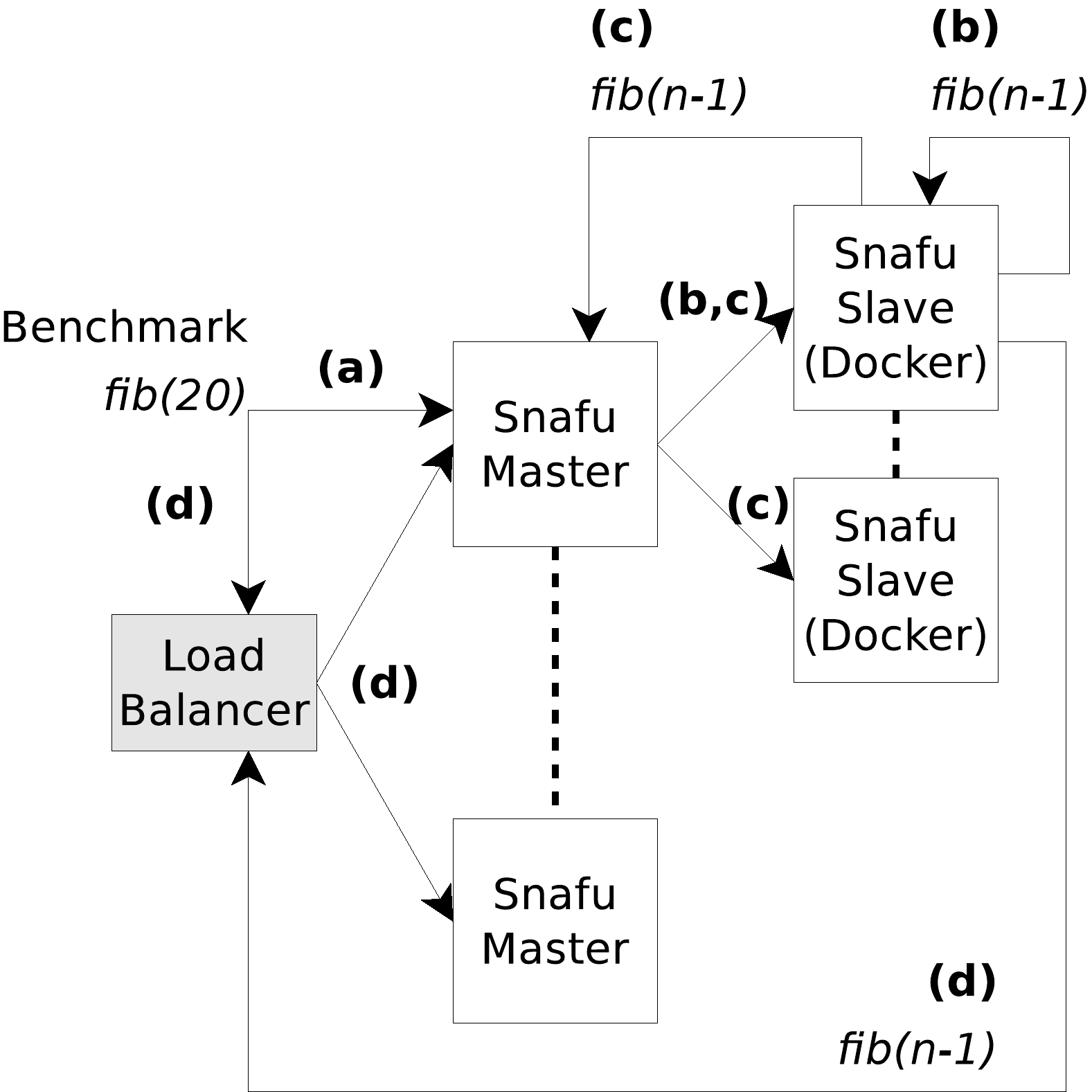}
\caption{Four benchmark setups to determine the performance of Snafu}
\label{fig:benchmark}
\end{figure}

The performance of a FaaS host depends not only on the function execution speed but also at the system topology
determined by scalability, isolation and authentication requirements.
Fig. \ref{fig:benchmark} combines four possible setups for operating Snafu in the cloud. Setup «a» is the
default configuration in which Python functions are executed in-process. The master-slave setup «b» forwards all function
requests to a containerised slave. All requests from functions to other functions, including recursive requests,
remain contained. Setup «c» forwards these requests back to the master instance which may decide to forward it back
again or to other container instances for load balancing. Setup «d» is the most complex one in which a load balancer
distributes load among multiple master instances. This paper reports on the setups «a»--«c».

Table \ref{tab:snafuperf} summarises the measured execution times of the recursive Fibonacci
function $fib$ implemented in Python according to setup «a». The function references a global counter variable which
tracks the number of function calls. On each invocation, the function calculates this number
times the sine of the number, leading to increasing execution times across calls for the non-isolated
executor. Each value is averaged over five calls with clean restarts after each measurement sequence.

\begin{table}[htb]
\centering
\caption{Snafu performance comparison in calls per second.\label{tab:snafuperf}}
\begin{tabular}{|l||r|r|r|} \hline
\textbf{Configuration}			& \textbf{fib(7)}	& \textbf{fib(12)}	& \textbf{fib(15)}	\\ \hline

Stateful in-process (IP)		& 81.91			& 298.90		& 327.28		\\ \hline
IP + debug output + CSV log		& 81.38			& 278.05		& 316.80		\\ \hline
IP + debug output + AWS4 authenticator	& 76.22			& 240.25		& 263.14		\\ \hline
Isolating in-process (IIP)		& 66.17			& 155.14		& 183.11		\\ \hline
IIP + AWS4 + debug + CSV log		& 62.78			& 146.56		& 162.96		\\ \hline\hline
External Python2 isolating executor	& 45.85			& 119.54		& 139.33		\\ \hline
External Python2 non-shared executor	& 7.22			& 7.77			& 7.79			\\ \hline\hline
External Docker shared executor		& 34.10			& 173.69		& 246.95		\\ \hline
Comparison: AWS Lambda			& 15.25			& 59.64			& 86.64			\\ \hline
\end{tabular}
\end{table}

The measurements clearly show that the performance is impaired by any configuration option desirable for production use.
However, not only the best, but also the worst in-process performance are still competitive compared to execution of the same
function in AWS Lambda with 377.75\% and 188.09\% relative speed for $fib(15)$, respectively.
Furthermore, the most closely matching configuration of isolated execution through containers
per tenant yield 223--285\% of Lambda's call performance.

Another result is the highly degraded performance when spawning external processes to execute functions. The degradation
is much smaller (57.42\% compared to 97.62\%) when re-using external interpreter instances;
the difference between isolated and stateful external execution
is a lot higher than between their stateful and stateless internal counterparts (50.21\%).
It should be
noted that the entire implementation is single-threaded apart from incoming request threads and future versions can thus be
expected to perform better when making use of automated parallelisation optimisation.

\begin{figure}[h]
\center
\includegraphics[width=0.493\columnwidth]{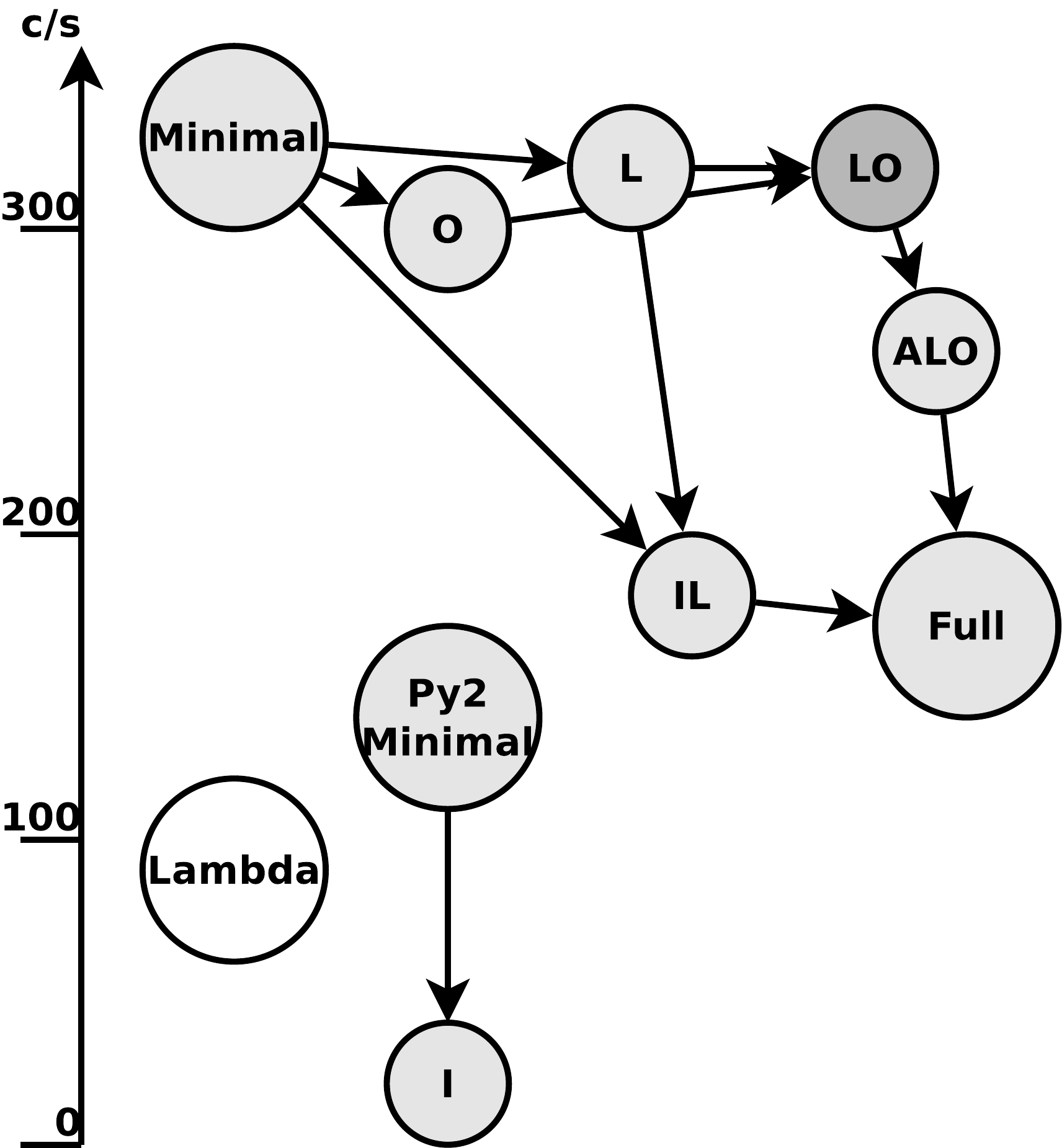}
\caption{Measured configuration space with the internal execution compared against external and AWS Lambda}
\label{fig:confspacemetrics}
\end{figure}

Fig. \ref{fig:confspacemetrics} picks up the configuration space from Fig. \ref{fig:confspace} again and relates
all configuration performance values based on the execution of $fib(15)$.

\subsection{Connection handling}

\begin{figure}[h]
\center
\includegraphics[width=0.49\columnwidth]{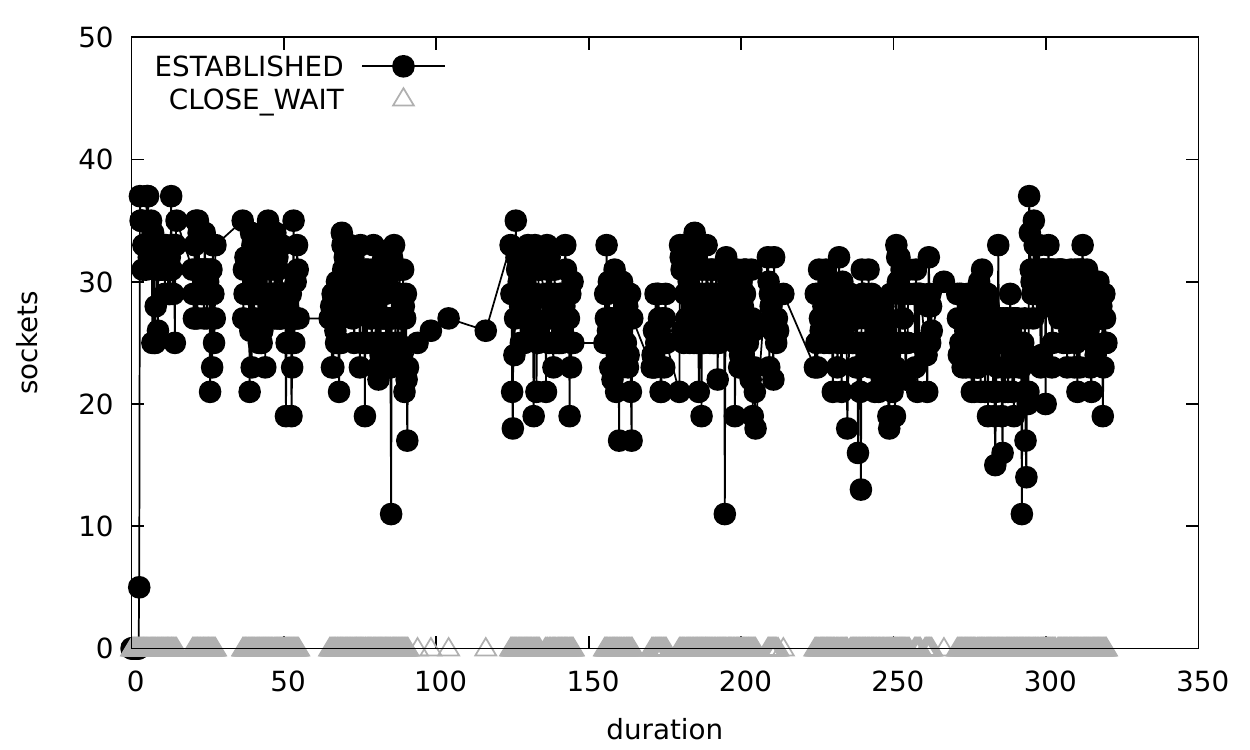}
\includegraphics[width=0.49\columnwidth]{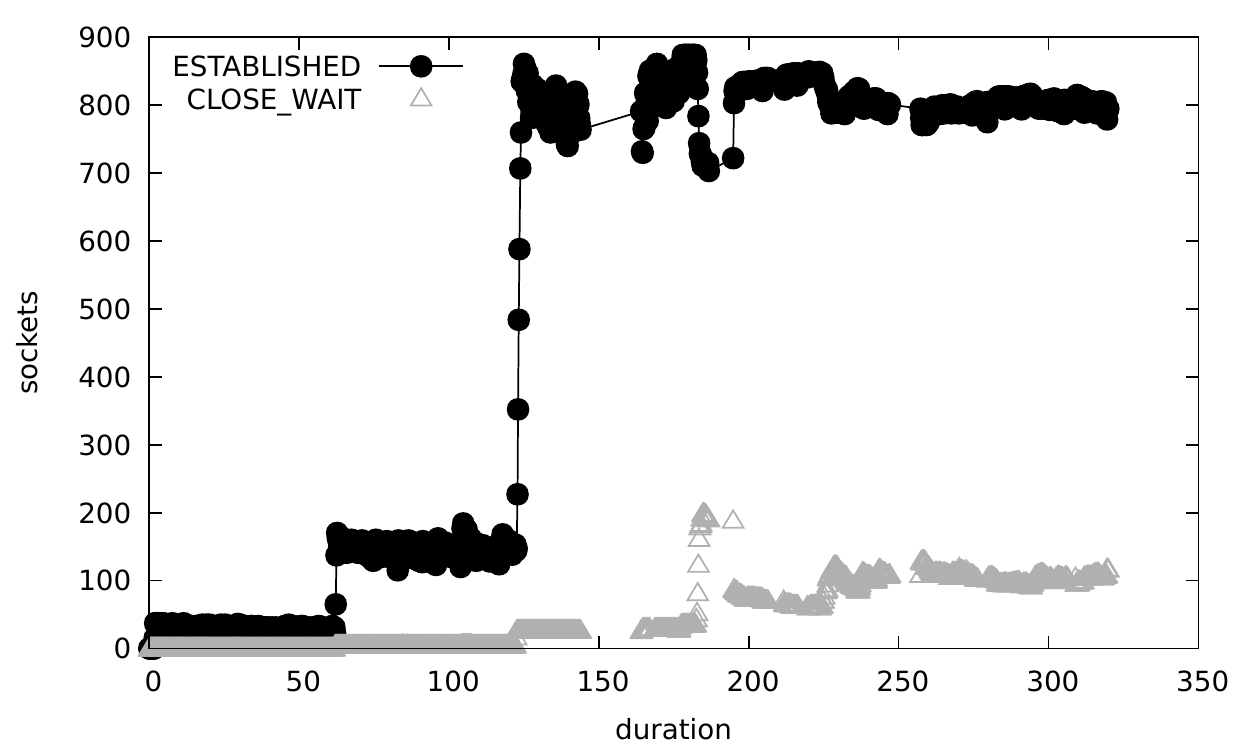}
\includegraphics[width=0.49\columnwidth]{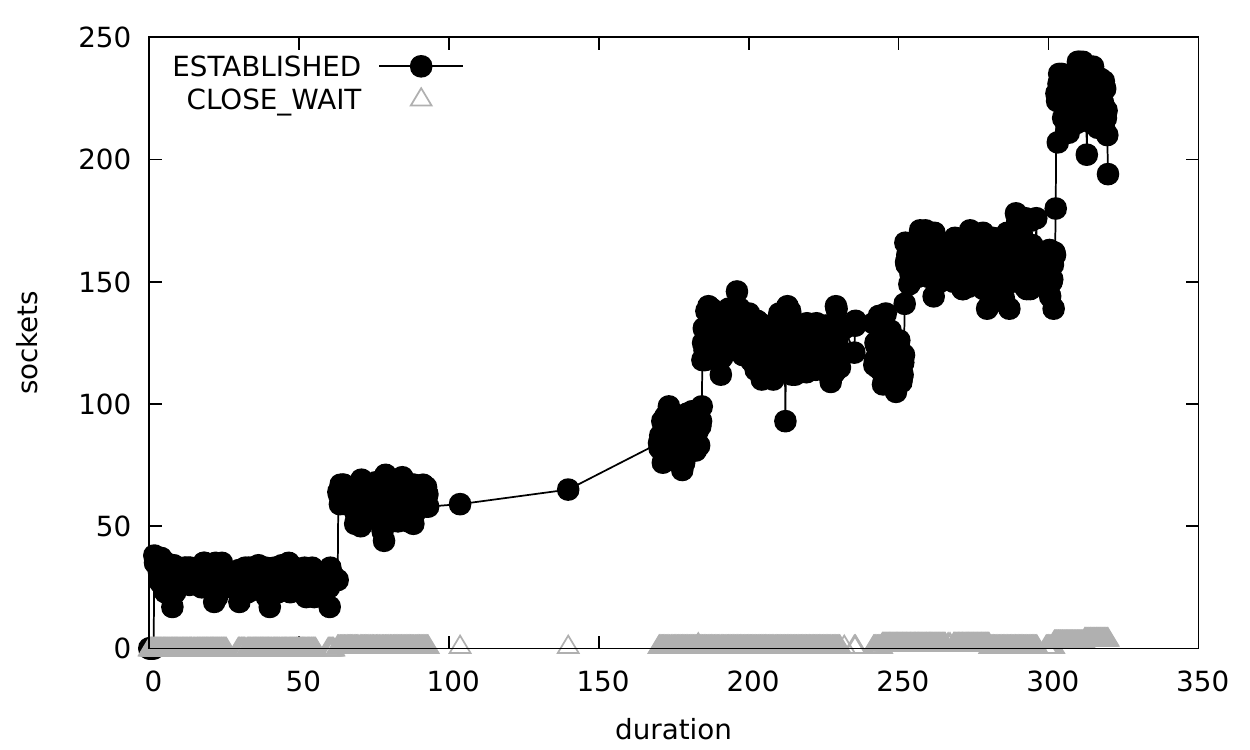}
\includegraphics[width=0.49\columnwidth]{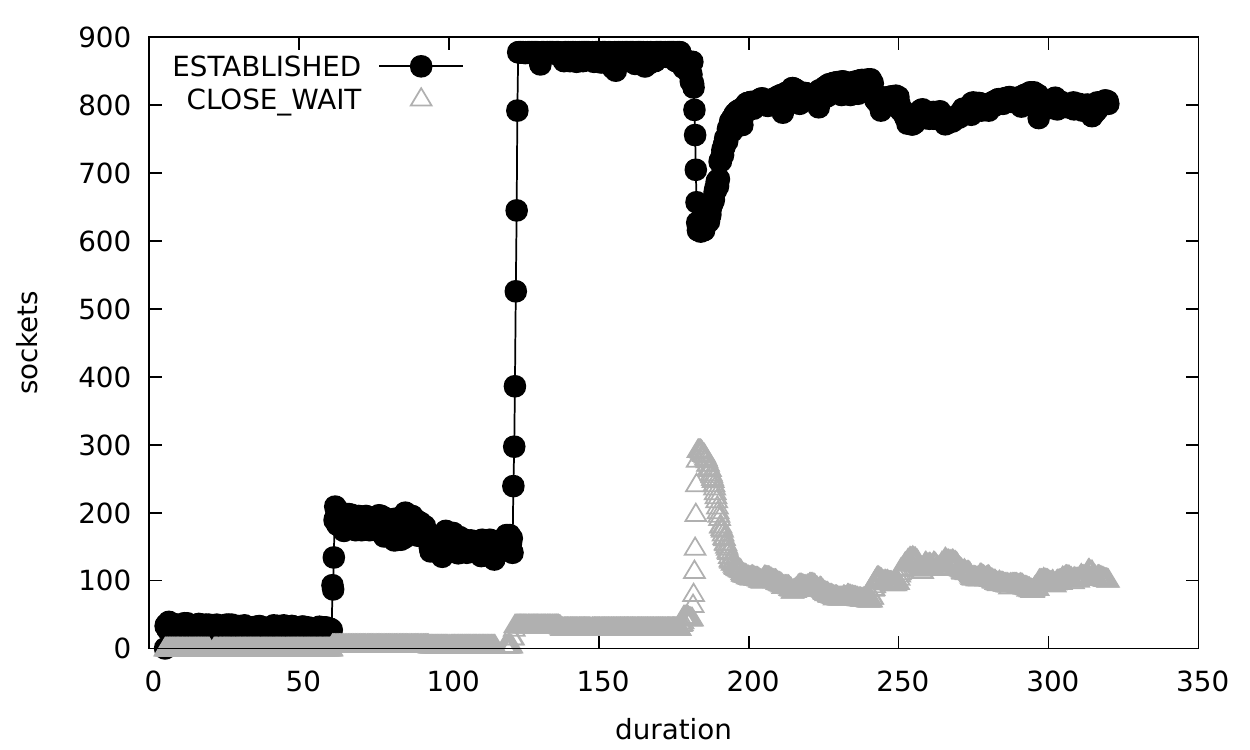}
\caption{Snafu open sockets over time while executing fib(25): all timeouts deactivated (upper left); AWS CLI timeouts deactivated (upper right); Boto timeouts deactivated (lower left); default timeouts (lower right) -- y scales not normalised}
\label{fig:sockets}
\end{figure}

A useful feature of Snafu is the socket reaper which works around inadequacies in the default invocation tool configuration provided by AWS and used
by many application engineers. These tools, both the AWS CLI and the Boto library for Python which is used for calling other functions
or for recursive calls from a function implemented in Python, have default socket read timeouts of 60 seconds each. Requests are repeated
when the timeout happens which implies that previous function instances keep running while new ones are added.
Fig. \ref{fig:sockets} compares the effect of the different configurations on the number of open sockets in Snafu, in particular
sockets in \texttt{CLOSE\_WAIT} state whose execution results are not read anymore by clients. The function semantics
are opaque, thus terminating function instances is not an option, but closing the lingering sockets is helpful to avoid a resource
exhaustion as most systems allow for only 1024 open file descriptors (including sockets) by default as determined by \texttt{ulimit -n}.

\begin{figure}[h]
\center
\includegraphics[width=0.7\columnwidth]{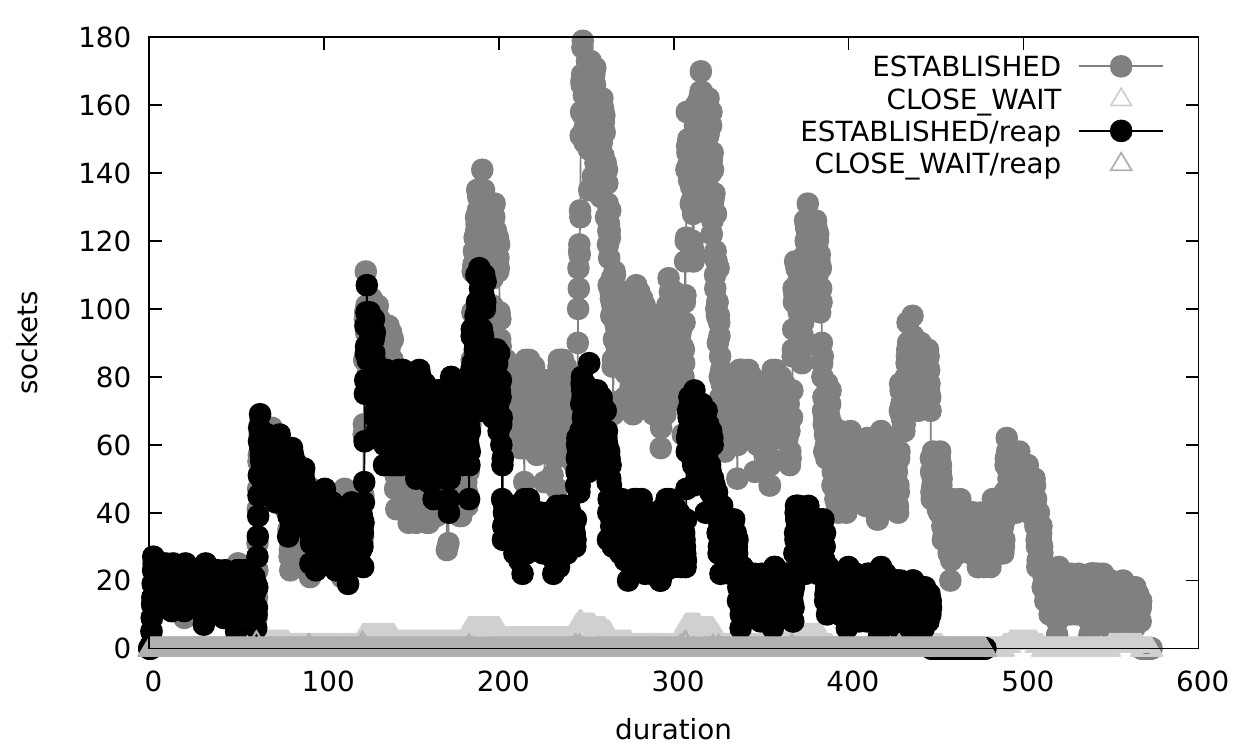}
\caption{Snafu open sockets over time while executing fib(15) after delay modification with and without reaper}
\label{fig:reaper}
\end{figure}

Fig. \ref{fig:reaper} shows the effect of using the reaper while keeping the AWS CLI and Boto default configuration.
The executed $fib$ function has been modified to wait a constant time of 0.1s in each invocation instead of performing the sine calculation. This modified function
can still be considered an edge case with worst processing to I/O ratio whereas other functions would give the reaper more
time to perform.
Using the reaper saves opening around 60 sockets, or one third, in the peak time, and shortens the process execution
considerably by around 120 seconds, or one fifth.

\subsection{Economics}

For a fair comparison of the cost of function invocations, Snafu needs to be operated in a setting comparable to
commercial FaaS providers. As AWS Lambda is currently the only provider supporting Python functions,
the experiments are conducted using the provider's Elastic Compute Cloud (EC2) which offers virtual machines
on demand.
The initial cost comparison experiment for setups «a»--«c» is conducted with an on-demand EC2 t2.small instance which albeit being
a bursty instance type is still expected to have a usable performance,
and a small Lambda instance with 128 MB of memory and a free tier of 1 million requests.
The calls per second ($cps$) are determined with a $fib(20)$ invocation which leads to 13529 recursive invocations.
Interpolating from the $cps$ metric, the upper bound for the calls per month ($cpm$) is determined
with $cpm = \frac{365}{12} \times 24 \times 60^2 \times cps$. Similarly, the price per month ($ppm$)
is calculated from the provider-specified price per hour ($pph$) with $ppm = \frac{365}{12} \times 24 \times pph$
or from the price per million calls ($ppmc$).

A utility index is defined to determine the performance in relation to the cost, calculated per the utility function
$\frac{cpm}{ppm \times 1000000}$. Table \ref{tab:snafucost} contains all results.
It becomes clear that the utility is highest for the unauthenticated in-process execution and lowest for the
always authenticated containerised execution. While the estimation was based on smaller measurement periods
which presumably benefited from burst performance, the overall observation is still that even in the worst case,
the utility of EC2-hosted Snafu for high-frequency invocation of the function under test is better than the one
achieved with AWS Lambda.

\begin{table}[htb]
\centering
\caption{Snafu economics comparison in cost per month, region us-west-1, currency CHF (1 CHF = $\approx$ 1.002 USD), function $fib(20)$.\label{tab:snafucost}}
\begin{tabular}{|l||r|r|r|r|r|} \hline
\textbf{Configuration}	& cps		& est. cpm		& base price	& est. ppm		& Utility	\\ \hline

AWS Lambda		& 99.30		& $\approx$ 260965583	& ppmc: 0.20	& $\approx$ 51.99	& 5.02		\\ \hline
Snafu IP «a»		& 347.19	& $\approx$ 912418507	& pph: 0.031	& $\approx$ 22.59	& 40.39		\\ \hline
Snafu Docker «b»	& 281.25	& $\approx$ 739126707	& -''-		& -''-			& 32.72		\\ \hline
Snafu Docker «c»	& 138.10	& $\approx$ 362928625	& -''-		& -''-			& 16.07		\\ \hline
\end{tabular}
\end{table}

\begin{figure}[h]
\center
\includegraphics[width=0.7\columnwidth]{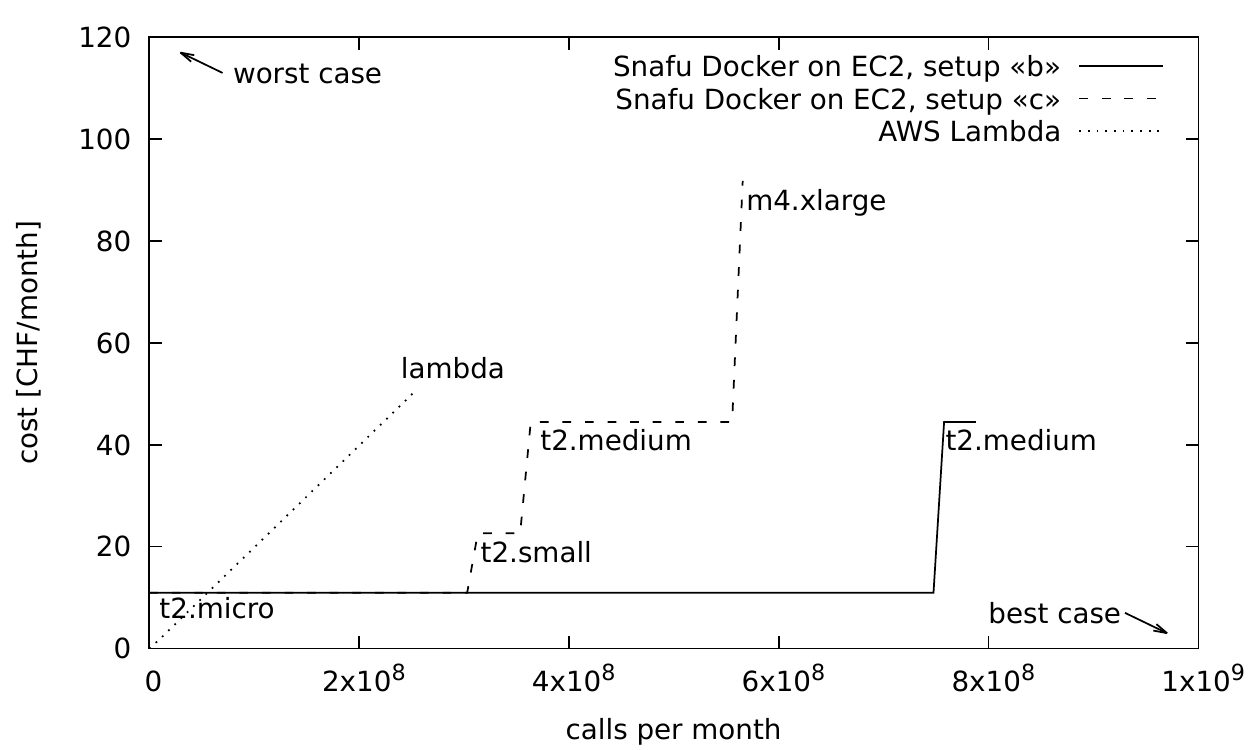}
\caption{Snafu and AWS Lambda performance and pricing comparison}
\label{fig:pricing}
\end{figure}

In practice, the choice of instance type needs to be clear before offering the service due to the inability of most
commercial cloud providers to scale instances vertically in fine-grained steps.
Fig. \ref{fig:pricing} explains which EC2 instance type is required depending on the setup («b» or «c») and the
expected calls per month. Due to the actual pay-per-use pricing model of AWS Lambda, it outperforms Snafu
for the benchmarked $fib(20)$ method for low use. For more frequent use (> 260 mio cpm), the Snafu deployments perform better and
are still more economical. For even heavier use (> 580 mio cpm), Snafu still performs with larger instance types but
quickly becomes prohibitively expensive. At some point (> 600 mio cpm), the internal scalability constraints of either
the FaaS host or the function implementation let the performance stuck while the price still climbs.

Recent research on microservice-based application architectures and cloud-native applications has focused
similarly on economical arguments. The authors of \cite{lambdacost} claim infrastructure cost reduction
of 70\% compared to an equivalent monolithic software when using AWS Lambda's Node.js executor for a very
specific call frequency. With Snafu, there is now the possibility to leave the application on EC2 but modularise
it on a fine-grained level and receive shared revenue from functions offered on cloud marketplaces.
The authors of \cite{microservicecosts} propose the CostHat model to determine the overall cost of both
monolithic and microservice-based applications for instance deployed on FaaS. Such cost models will
evolve to cover more flexible and requirements-driven deployments of functions whose pricing changes according
to per-tenant and per-function multipliers for scalability, isolation and authentication, among other factors.

\section{Conclusion}

The design space for FaaS hosts is still largely empty despite an increasing number of proposed and
implemented systems. Snafu contributes a distinct design which does not overlap with any of the
existing hosts. Its architecture is modular and extensible, and its implementation is competitive
with existing commercial and open source FaaS runtimes.
In particular, the flexible design makes it possible to balance isolation and authentication with
raw performance speed which sets it apart from comparable systems.
Snafu thus contributes to the broad availability of usable FaaS tools and therefore to the
feasibility of engineering applications on top of this service class.
Cloud platform providers benefit from operating Snafu in multi-tenant mode, cloud application
engineers avail themselves of a low-effort tooling to test locally without incurring cost,
and academics get a modular system to study and extend.
Limitations to overcome in the future include the lack of an asynchronous function model and
a performance prediction and tuning component.

\section*{Repeatability}

Snafu is publicly available at \url{https://github.com/serviceprototypinglab/snafu}.
Implementations of all referenced functions as well as invocation instructions are provided for reproducing
the stated results in the project wiki \url{https://github.com/serviceprototypinglab/snafu/wiki} which
serves as continuous scientific open notebook.

\section*{Acknowledgements}

This research has been supported by an AWS in Education Research Grant which helped us to run our experiments on AWS Lambda as representative public commercial FaaS.

\bibliographystyle{unsrt}
\bibliography{snafu}

\begin{thebibliography}{10}

\bibitem{microservicesscale}
Nicola Dragoni, Ivan Lanese, Stephan~Thordal Larsen, Manuel Mazzara, Ruslan
  Mustafin, and Larisa Safina.
\newblock {Microservices: How To Make Your Application Scale}.
\newblock CoRR abs/1702.07149, February 2017.

\bibitem{chatbot}
Mengting Yan, Paul~C. Castro, Perry Cheng, and Vatche Ishakian.
\newblock {Building a Chatbot with Serverless Computing}.
\newblock In {\em 1st International Workshop on Mashups of Things and APIs,
  MOTA@Middleware}, pages 5:1--5:4, Trento, Italy, December 2016.

\bibitem{dripcast}
Ikuo Nakagawa, Masahiro Hiji, and Hiroshi Esaki.
\newblock {Dripcast - Architecture and Implementation of Server-less Java
  Programming Framework for Billions of IoT Devices}.
\newblock {\em Journal of Information Processing (JIP)}, 23(4):458--464, 2015.

\bibitem{serverlessworkflows}
Maciej Malawski.
\newblock {Towards Serverless Execution of Scientific Workflows -- HyperFlow
  Case Study}.
\newblock In {\em 11th Workshop on Workflows in Support of Large-Scale Science
  (WORKS@SC)}, volume CEUR-WS 1800 of {\em CEUR Workshop Proceedings}, pages
  25--33, Salt Lake City, Utah, USA, November 2016.

\bibitem{serverless}
Scott Hendrickson, Stephen Sturdevant, Tyler Harter, Venkateshwaran
  Venkataramani, Andrea~C. Arpaci-Dusseau, and Remzi~H. Arpaci-Dusseau.
\newblock {Serverless Computation with OpenLambda}.
\newblock In {\em 8th USENIX Workshop on Hot Topics in Cloud Computing
  (HotCloud)}, Denver, Colorado, USA, June 2016.

\bibitem{pywren}
Eric Jonas, Shivaram Venkataraman, Ion Stoica, and Benjamin Recht.
\newblock {Occupy the Cloud: Distributed Computing for the 99\%}.
\newblock preprint at ar$\chi$iv:1702.04024, February 2017.

\bibitem{cloudcontrolplane}
Josef Spillner.
\newblock {Exploiting the Cloud Control Plane for Fun and Profit}.
\newblock preprint at ar$\chi$iv:1701.05945, January 2017.

\bibitem{cloudcomplexity}
Michel Catan, Roberto~Di Cosmo, Antoine Eiche, Tudor~A. Lascu, Michael
  Lienhardt, Jacopo Mauro, Ralf Treinen, Stefano Zacchiroli, Gianluigi
  Zavattaro, and Jakub Zwolakowski.
\newblock {Aeolus: Mastering the Complexity of Cloud Application Deployment}.
\newblock In {\em Service-Oriented and Cloud Computing -- Second European
  Conference (ESOCC)}, pages 1--3, Málaga, Spain, September 2013.

\bibitem{lambdacost}
Mario Villamizar, Oscar Garces, Lina Ochoa, Harold~E. Castro, Lorena Salamanca,
  Mauricio Verano, Rubby Casallas, Santiago Gil, Carlos Valencia, Angee
  Zambrano, and Mery Lang.
\newblock {Infrastructure Cost Comparison of Running Web Applications in the
  Cloud Using AWS Lambda and Monolithic and Microservice Architectures}.
\newblock In {\em 16th IEEE/ACM International Symposium on Cluster, Cloud and
  Grid Computing (CCGrid)}, pages 179--182, 2016.

\bibitem{microservicecosts}
Philipp Leitner, Jürgen Cito, and Emanuel Stöckli.
\newblock {Modelling and Managing Deployment Costs of Microservice-Based Cloud
  Applications}.
\newblock In {\em 9th IEEE/ACM International Conference on Utility and Cloud
  Computing (UCC)}, Shanghai, China, December 2016.

\end{thebibliography}

\end{document}